\documentclass[rmp,twocolumn,showpacs,superscriptaddress]{revtex4}

\usepackage{graphicx,amsmath,amssymb,dcolumn,txfonts}

\begin{document}

\title{Comment on ``Regularizing capacity of metabolic networks''}

\author{Petter Holme}
\affiliation{Department of Computer Science, University of New Mexico,
  Albuquerque, NM 87131, U.S.A.}

\author{Mikael Huss}
\affiliation{School of Computer Science and Communication,
  Royal Institute of Technology, 10044 Stockholm, Sweden}

\begin{abstract}
In a recent paper, Marr, M\"{u}ller-Linow and H\"{u}tt [Phys.\ Rev.\ E
\textbf{75}, 041917 (2007)] investigate an artificial dynamic system on
metabolic networks. They find a less complex time evolution of this
dynamic system in real networks, compared to networks of null
models. The authors argue that this suggests that metabolic network
structure is a major factor behind the stability of biochemical steady
states. We reanalyze the same kind of data using a dynamic system
modeling actual reaction kinetics. The conclusions about stability,
from our analysis, are inconsistent with those of Marr~\textit{et al.}
We argue that this issue calls for a more detailed type of modeling.
\end{abstract}


\maketitle

\section{Introduction}

Within living organisms, matter is constantly converted between
different molecular species. It is often assumed that concentrations
of metabolites tend to settle into steady states (rather than showing
periodic or chaotic behavior)~\cite{giersch:control}. Furthermore,
experimental studies of metabolic pathways are typically performed
under steady-state conditions~\cite{fell:book}. Although the steady
state is, strictly speaking, a mathematical abstraction, it is
nevertheless a useful reference state~\cite{hofmeyr:mini}.
The steady-state assumption is also fundamental to traditional
metabolic control analysis~\cite{fell:book}.
In a recent article~\cite{marr:regula} Marr, M\"{u}ller-Linow and
H\"{u}tt hypothesize that the network structure of metabolism is an
important factor for promoting a steady state, rather than a complex,
dynamics. The authors run an artificial dynamic system on the
networks. In short, a vertex $i$ can have two states, $0$ or $1$. If
the sum of the state variables in the neighborhood of $i$ exceeds a
fixed threshold, then $i$ changes to the other state. This scheme is
very different from real biochemical dynamics and traditional models
of reaction kinetics. First, the state variables in metabolic models
are usually continuous (concentrations~\cite{fell:book}) or sometimes
discrete variables (molecule counts~\cite{kattunge}), but never
binary. Second, the sum of these variables is conserved (if in-
and outflow is neglected); whereas, in the dynamics of Marr \textit{et
  al.}, this is not the case. This dynamics does in fact not reach a
steady state; instead, the authors analyze the complexity of the
output time series with entropy-like measures. They conclude that real
metabolic networks give a less complex output than different ensembles
of model networks, and argue that this implies that the structure of
real metabolic networks may promote steady state dynamics. However, no
theory for how the values of the ``entropies'' of the binary dynamics
relate to the stability of metabolic flux is given. Many
authors have used binary dynamics to model other processes of cellular
biology, such as signal transduction (e.g.\ Ref.~\cite{brandman}) or
genetic regulation (e.g.\ Ref.~\cite{kauffman:glass}). However, these
models explicitly try to describe systems in which the components, at
least under some circumstances, show switch-like, binary
behavior~\cite{brandman}. Moreover, the lack of detailed understanding
of e.g.\ gene expression makes it more natural to use simplified
dynamics to uncover underlying principles. Metabolic reactions follow
known principles of chemistry and are described by well-established
differential equation models. Therefore, in contrast to the mentioned
information processes, there is no need for, or natural interpretation
of, a binary description of the metabolism. Therefore the
conclusions of Marr \textit{et al.}\ can be validated by comparing
their results to results obtained using such standard differential equation
modeling. We follow this approach to find no significant difference
between real and null-model network. Our simulations in this Comment are
rather limited, but sufficient to make the point that Marr \textit{et
  al.}'s study cannot rule out the possibility that stable
steady-states can be explained more simply, as direct consequences of
the reaction kinetics, rather than as a result from the interaction
between the network topology and the dynamic system.

\begin{figure}
  \centering\resizebox*{\linewidth}{!}{\includegraphics{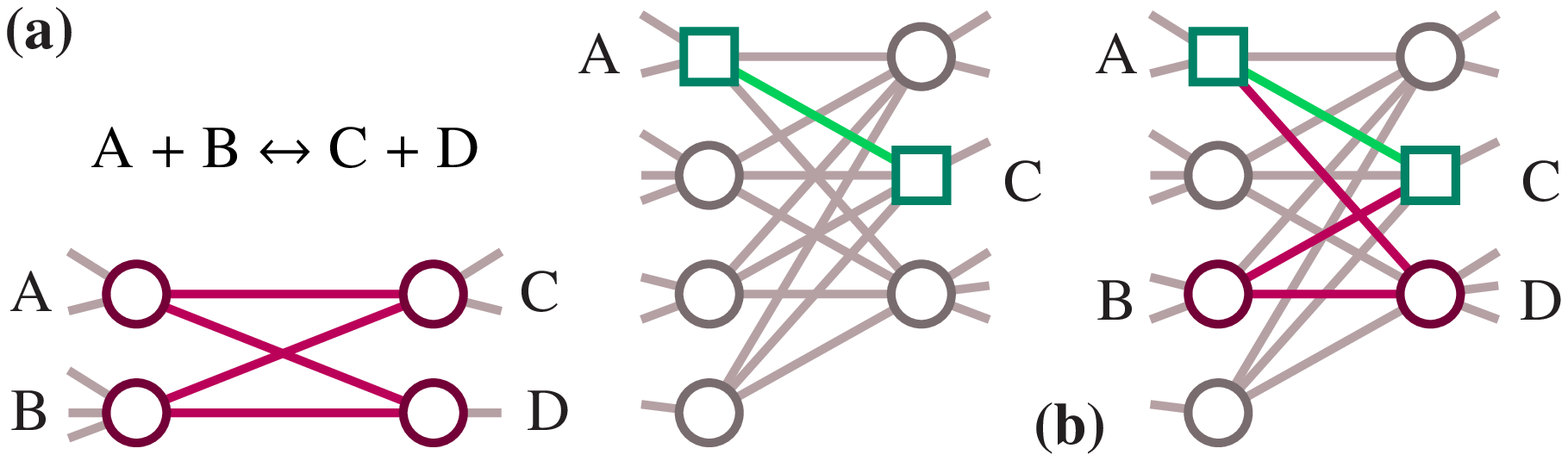}}
  \caption{(Color online.) Illustration of the construction of substrate graphs from
    chemical reactions (a), and our method for recreating a reaction
    system from a substrate graph (b).
  }
  \label{fig:ill}
\end{figure}

\section{Constructing a reaction system}

Marr \textit{et al.}'s examples of real metabolic networks are taken
from a work of Ma and Zeng (MZ)~\cite{ma:zeng}. They are
\textit{substrate graphs}~\cite{zhao:meta} where a substrate is linked
to the products of a reaction (Fig~\ref{fig:ill}(a)). MZ also
preprocessed the data by omitting ubiquitous ``currency
metabolites''~\cite{our:bio}. We also use the MZ networks as the starting point for our simulations. In constructing the networks,
information is lost; so if one wants to simulate the reaction system
with a substrate graph as starting point, one needs (explicitly or
implicitly) to recreate the
reactions via a model. From two assumptions about reaction systems, we
propose a simple scheme to create a plausible set of reactions that
can be reduced to a given substrate graph. We first assume that all
reactions are of a \textit{2--2-form}: A + B $\leftrightarrow$ C + D,
or \textit{2--1-form}: A + B $\leftrightarrow$ C. These are the most
common forms of biochemical reactions. We do not include more complex
reactions, mostly because it would significantly increase the
computational complexity. Our second assumption is that the number of
reactions creating an edge in the substrate graph is rather small (we
confirm \textit{a posteriori} that the average number of reactions per
vertex is similar to that of the real substrate graph). The following
algorithm is a simple way of fulfilling these assumptions:
\begin{enumerate}
\item Start with all edges unmarked.
\item \label{step:pick} Pick one unmarked edge $(\mathrm{A},\mathrm{C})$.
\item \label{step:bip} Find the maximal, full bipartite subgraph
  $K_{(\mathrm{A},\mathrm{C})}$ (a subgraph consisting of two vertex
  sets, and edges between every pair of vertices in the different sets,
  but no edge between vertices in the same set) that contains
  $(\mathrm{A},\mathrm{C})$. (See Fig.~\ref{fig:ill}(b).)
\item Pick one four-cycle of $K_{(\mathrm{A},\mathrm{C})}$ including
  $(\mathrm{A},\mathrm{C})$ (say  $(\mathrm{A},\mathrm{C},\mathrm{B},
  \mathrm{D},\mathrm{A})$).  (See Fig.~\ref{fig:ill}(b).)
\item \label{step:add} Add A + B $\leftrightarrow$ C + D to the set of
  reactions and mark the edges $(\mathrm{A},\mathrm{C})$,
  $(\mathrm{C},\mathrm{B})$, $(\mathrm{B},\mathrm{D})$,
  $(\mathrm{D},\mathrm{A})$.
\item If there are unmarked edges go to step~\ref{step:pick}.
\end{enumerate}
$K_{(\mathrm{A},\mathrm{C})}$ gives all reactions of the 2--2-form
that induce the edge $(\mathrm{A},\mathrm{C})$ in a substrate
graph. If $K_{(\mathrm{A},\mathrm{C})}$ is empty at
step~\ref{step:bip}, then $(\mathrm{A},\mathrm{C})$ must have been
generated by a reaction of a 2--1-form. Therefore, instead of
$K_{(\mathrm{A},\mathrm{C})}$, consider the set $L_{(\mathrm{A},\mathrm{C})}$, of
$(\mathrm{A},\mathrm{C})$ and its adjacent edges, and deduce reactions
of the 2--1-form at step~\ref{step:add}.

\section{Reaction kinetics}

Once we have a set of reactions, derived from a substrate graph, we simulate the biochemical dynamics
by standard Michaelis--Menten (MM) kinetics~\cite{gutfreund:kin}
supplemented by noise in the enzyme and substance concentrations. MM
kinetics are used to model enzyme-catalyzed reactions, which are
common in metabolism. The MM description builds on the principle of
mass-action~\cite{gutfreund:kin}---the forward rate of a
reaction is proportional to the product of the concentrations of the
substrates---and uses additional assumptions to simplify the resulting
rate laws. For the reaction $i$, A + B $\rightarrow$ C + D (which
is assumed to be enzyme-catalyzed), we use a version of the
two-substrate MM rate law to calculate the flux
\begin{equation} \label{eq:rate}
  r_i = \frac{E_i+\eta_E}{
    k_0^{-1}+1/k_1 [\mathrm{A}] +1/k_2 [\mathrm{B}] +
    1/k_{12}[\mathrm{A}][\mathrm{B}]}~.
\end{equation}
where $\eta_E$ is a random variable modeling fluctuations in the
enzyme concentration. The time evolution of the concentration of a
substance A is then determined by
\begin{equation} \label{eq:mac}
  \frac{\mathrm{d}[\mathrm{A}]}{\mathrm{d}t}= \eta_\mathrm{conc.} + 
  \sum_i\pm r_i ,
\end{equation}
where the sum is over all A's reactions, the sign in front of $r_i$ is
positive (negative) if A is a
product (substrate) of $i$, and $\eta_\mathrm{conc.}$ is a random
noise term modeling fluctuations due to in- and outflow of A. We seek
to use the same information as in Ref.~\cite{marr:regula}, therefore
we do not use empirical parameter values (which are, anyway, hard to
obtain). Instead, we assign parameters in arbitrary units, but we
choose them to be reasonable relative to one another. $\eta_E$ and
$\eta_\mathrm{conc.}$ are normally distributed $N(0,0.002)$ (the first
argument is the mean; the second is the standard deviation). New
values for $\eta_E$ and $\eta_\mathrm{conc.}$ are drawn every time step of
the integration.  The initial values of substance concentrations, enzyme
concentrations and reaction coefficients ($k_0$, $k_1$, $k_2$ and
$k_{12}$) are drawn from $N(1,1)$, $N(0.2,0.2)$ and $N(0.1,0.1)$
respectively.  We choose $\mu_k=1\times 10^{-3}$, $\sigma_k=5\times
10^{-4}$, $\mu=1$ and $\sigma=0.5$. The equations are integrated with
a second order Runge--Kutta--Helfand--Greenside scheme with time step
0.1, total running time 2500 time units, and 20 averages over
different sets of initial configurations. These runs are, for each
network, averaged over 20 realizations of the reaction-system
construction. For comparison, we also run the dynamics on 100 samples
of one Marr \textit{et al.}'s null-model networks---random graphs with
the same degree sequence as the original network. These are obtained
by random rewiring---we go through all edges and for each edge $(i,j)$
randomly pick an edge $(i',j')$ and replace these edges with $(i,j')$
and $(i',j)$ (unless this would introduce a multiple edge or a
self-edge, in which case a new random edge $(i',j')$ is selected).

\begin{figure}
  \centering\resizebox*{0.9 \linewidth}{!}{\includegraphics{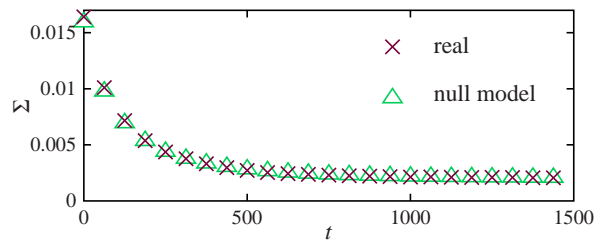}}
  \caption{(Color online.) Time evolution of the average standard
    deviation $\Sigma$ of the flux. Standard errors are smaller than
    the symbol size.
  }
  \label{fig:dyn}
\end{figure}

\section{Simulation results}

In the simulations, a vast majority of the substance concentrations
converge to steady states. A few subnetworks of oscillating or chaotic
concentrations may exist~\cite{chaos:moreno}, but in this Comment we
focus on bulk properties. To study the approach to
equilibrium, and fluctuations, we measure the average standard
deviation of the flux through the substances,
\begin{subequations}
\begin{eqnarray}
  \Sigma & = & \sqrt{\frac{1}{n} \sum_i \Phi_i^2 -
    \Big[\frac{1}{n}\sum_i\Phi_i\Big]^2}, \mbox{~where}\label{eq:sigma}\\
  \Phi_i & = & \frac{1}{2}\sum_j |r_j|.\label{eq:phi}
\end{eqnarray}
\end{subequations}
The sums in Eq.~\ref{eq:sigma} are over all substances; the
sum in Eq.~\ref{eq:phi} is over $i$'s reactions; the factor $1/2$
comes from the double count of mass-flow in and out of substance $i$,
and  $n$ is the number of substances.
In a plot
of $\Sigma(t)$, the stability of the steady state can be monitored by
two quantities. First, when the system approaches an equilibrium,
$\Sigma(t)$ will decrease---a faster decrease implies a more stable
system. Second, a lower equilibrium level means that the system has
less cyclic or chaotic components, and responds faster to
perturbations from the noise, and therefore has more stable steady
states. In Fig.~\ref{fig:dyn} we plot $\Sigma(t)$ for MZ's human
metabolic network and the null-model networks. The two curves almost
overlap. For different null-model realizations the curves may deviate
slightly from those of the real network, but there the
null model cannot be rejected with any high level of significance. The
equilibrium level is of the order of the noise, which means that
periodic and chaotic behavior is almost fully suppressed. To be more
systematic, we note that $\Sigma(t)$ display two aspects of
stability---how fast equilibrium is reached and the height of the
equilibrium level.  Since the
curves do not fit any simple functional form we measure the half-time
(the time to reach midway between $\Sigma(0)$ and $\Sigma(\infty)$)
using spline interpolation. The p-value (fraction of null-model
observations lower than the real value) of the half-time is 67\%; the
corresponding value of the equilibrium level is 65\%. The effect of
the difference in network structure between the real and null-model
networks is thus, in this case, negligible. We test a few other sets
of parameter values, organisms (\textit{E. coli} and
\textit{M. musculus}), and a more straightforward mass action
kinetics; and, in all cases, arrive at the above conclusion. A full
scan of the parameter space would be interesting, however, in this
Comment we just make the point that the conclusion of  Marr \textit{et
  al.}\ can be inconsistent with more realistic simulations, and leave
the insignificance of network structure to the stability of metabolic
steady states as a conjecture.

\section{Conclusions}

To conclude; simple, stylized dynamic systems---``dynamic
probes''---are valuable tools for studying complex biological systems.
We believe that these should be designed to model the real dynamics as
closely as possible. Marr \textit{et al.}~\cite{marr:regula} propose a
dynamic probe to study metabolic steady states that violates many of
the known features of reaction kinetics; it does not even reach a
steady state---the objective of the study as given in their Abstract.
By the collective effort of researchers, our understanding of the
metabolism continuously advances. This does, however, not improve the
approach of Marr \textit{et al.}\ as their dynamics does not make use
of biochemical information. Relative to biological
information processes, metabolism is a rather simple system, and is
believed to be well-described by simple differential equation
models. For the simulations we carry out, the null-model networks are
not less efficient than the real networks in suppressing complex
dynamics. If this holds in general, then steady-state stability is
a fundamental property of chemical reaction systems. The study of Marr
\textit{et al.}\ does not rule out such a simple explanation of
steady-state behavior---an explanation that is a common opinion in
biochemical literature (cf.\ Hofmeyr and
Cornish-Bowden's dictum ``mass-action is the intrinsic driving force
for self-organization of reaction networks''~\cite{hofmeyr:mini}).
If the reaction kinetics is the sole cause of metabolic steady-states,
the steady-state dynamics is a constraint to, rather than an outcome
of, natural selection. This situation is reminiscent of the power-law
degree distribution of metabolic networks. Such distributions can also
be seen in astrochemical networks that are not subject to natural
selection~\cite{wagner:robu}. We believe the question of dynamic
stability in metabolism should be studied at a more detailed level
than networks, avoiding the reduction of database information to
substrate graphs. This is not to say that graph theory is useless in
the study of metabolism. On a large scale, metabolic networks are
different from the null-model networks. This is a valid conclusion
from Ref.~\cite{marr:regula}---the authors manage to separate the real
metabolic networks from both the null-model network we use, and
various other types of model networks derived from the original
graph. We believe much information about the organization of
metabolism lies in the answer to how this separation occurs. On
smaller scales, network theory can be used to find e.g.\ functional
modules~\cite{our:bio} and functions of individual
metabolites~\cite{gui:meta}.

~

\noindent\fbox{\begin{minipage}{0.96\columnwidth}
In a reply~\cite{marr2} to this comment the Marr
  \textit{et al.}\ argue that (with reference to our reconstruction of
  the reaction system) the fact that reaction coefficients are larger
  in one direction effectively break the network into disconnected
  components with respect to the dynamics. This is, of course, true in
  real reaction systems too, and a possible stabilizing mechanism for
  metabolism. However, it is irrelevant to our point that Marr
  \textit{et al.}, in Ref.~\cite{marr:regula}, by measuring an
  ``entropy'' of a time series from a dynamic system
  unrelated to reaction kinetics, cannot draw conclusions about the effect
  of metabolic network structure on biochemical dynamics. To make this
  point, we use the same starting point as Marr \textit{et al.}, it
  should be easy to check that our conclusions also hold for the MM
  dynamics on full reaction systems from metabolic databases.
\end{minipage}
}

\acknowledgments{
  PH acknowledges financial support from Wenner--Gren
  Foundations and National Science Foundation (grant
  CCR--0331580). The authors are grateful for comments from Jing Zhao
  and Martin Rosvall.
}

\end{document}